\documentclass[twocolumn]{article}
\usepackage[
    colorlinks=true,
    linkcolor=blue,
    citecolor=blue,
    urlcolor=blue,
    breaklinks=true
]{hyperref}
\usepackage[
  top=1.6cm,
  bottom=1.6cm,
  left=1.6cm,
  right=1.6cm,
  columnsep=0.5cm
]{geometry}
\usepackage{amsmath}
\usepackage{booktabs}
\usepackage{orcidlink}
\usepackage{graphicx,natbib}
\usepackage{authblk}
\usepackage{sectsty}
\usepackage[normalem]{ulem}
\usepackage{xcolor}
\usepackage{subfigure}
\usepackage{tabularx}
\begin{document}

\title{Observations of Binary Stars with the 1.3-m Devasthal Fast Optical Telescope Using Speckle Interferometry: An Attempt}

\author{Km Nitu Rai\thanks{E-mail: niturai201296@gmail.com}\orcidlink{0000-0002-9939-805X}}
\affil[1]{\small{\em{Aryabhatta Research Institute of Observational Sciences, Manora Peak, Nainital 263129, India.}}}
	
\author[1]{Arjun Dawn}	

\author[1]{Neelam Panwar\orcidlink{0000-0002-0151-2361}}

\author[1]{Jeewan C Pandey\orcidlink{0000-0002-4331-1867}}

\author[2,3]{Subrata Sarangi\orcidlink{0000-0001-6500-9987}}
\affil[2]{\small{\em{School of Applied Sciences, Centurion University of Technology and Management, Odisha-752050, India.}}}
\affil[3]{\small{\em{Visiting Associate, Inter-University Centre for Astronomy and Astrophysics, Post Bag 4, Ganeshkhind, Pune 411 007, Maharashtra, India.}}}

\author[4]{Prasenjit Saha\orcidlink{0000-0003-0136-2153}}
\affil[4]{\small{\em{Physik-Institut, University of Zurich,	Winterthurerstrasse 190, 8057 Zurich, Switzerland.}}}

\setbox0=\vbox{\hsize=7.1truein
  \hbox to \hsize{\hss\bf Abstract\hss}
  \smallskip \noindent 

We present a feasibility study of implementing optical interferometry and speckle techniques with the 1.3-m Devasthal Fast Optical Telescope (DFOT) at Aryabhatta Research Institute of Observational Sciences (ARIES), which is currently dedicated to photometric observations. Using the sCMOS camera at the DFOT backend, we perform interferometric speckle observations of six binary stars. Standard Speckle Interferometry algorithms are applied to analyze the recorded speckle images. While this study does not aim to achieve the diffraction limit of DFOT or address a full science-driven resolution case, it serves as a crucial testbed for instrumentation, data acquisition, and analysis of Speckles with DFOT. Although the individual speckle patterns exhibit significant positional variations from frame to frame, the Speckle Interferometry analysis successfully recovers the characteristic three-peak structure of the wide binaries, allowing their relative separations and position angles to be determined, demonstrating the viability of the approach. The measured angular relative separations of $\gamma$ Leo, $\gamma$ Vir, and $\zeta$ Her are $4.744'' \pm 0.012''$, $3.387'' \pm 0.027''$, and $1.582'' \pm 0.026''$, respectively, with corresponding projected position angles of $126.94^{\circ} \pm 0.14^\circ$, $171.57^{\circ} \pm 0.33^\circ$, and $81.06^{\circ} \pm 0.91^\circ$. These measurements are consistent with previously reported values in the literature, which provides strong motivation for more systematic observations and future implementation of optical interferometry techniques at meter-class telescopes.}
\date{\box0}

\maketitle


\section{Introduction}\label{sec:intro}

Interferometry is an imaging technique that uses the interference produced from the superposition of coherent signals to achieve high angular resolution. In the twentieth century, high-angular-resolution optical astronomy advanced largely through three distinct methods: Michelson stellar interferometry \citep[MSI;][]{michelson-Pease-1921}, Hanbury Brown-Twiss or intensity interferometry \citep[II;][]{HBT56,hanbury1974}, and speckle interferometry \citep[SI;][]{labeyrie1970}. Though all three use starlight correlations to obtain spatial information beyond conventional seeing-limited imaging, their underlying physics and instrumental setups are fundamentally unique. 

The MSI combines light collected by spatially separated telescopes to measure first-order coherence through interference of the optical fields. The technique requires accurate control of the optical path difference and is sensitive to atmospheric phase fluctuations, particularly for long-baseline observations. The II utilizes a fundamentally different approach, measuring second-order correlations in intensity fluctuations between spatially separated telescopes without requiring direct optical beam combination or phase stability. Although this makes II less sensitive to atmospheric phase fluctuations and enables the use of long baselines, its sensitivity is limited by photon statistics, and it primarily provides visibility-amplitude information \citep{Kieda22, astri2023}.
The SI is a single-aperture technique in which atmospheric turbulence is effectively frozen by recording a sequence of short-exposure images. Statistical analysis of these speckle patterns, typically using Fourier transform or autocorrelation techniques, enables the extraction of spatial information at angular resolutions approaching the telescope's diffraction limit.
Thus, SI exploits short-exposure temporal sampling and post-observation processing to recover spatial information lost in conventional long-exposure imaging. It provides a relatively simple and cost-effective alternative to the more complex and expensive instrumentation required for MSI and II, while having its own limitations in accuracy, sensitivity, and applicability.

Over the past few decades, SI has matured into a powerful technique for achieving diffraction-limited resolution with ground-based telescopes. Since its early demonstration \citep{labeyrie1970}, SI has been successfully applied worldwide to a broad range of science cases, including the detection and orbital characterization of binary and multiple stellar systems and the discovery of new companions \citep{mcalister1977speckle, foy1985multiple, hartkopf1999iccd, horch2017, Howell_2025}. Early applications of speckle detectors in the late 1970s and 1980s, including work by \citet{breckinridge1979kitt, bonneau1980resolution, prieur1998pisco}, established SI as a practical technique for resolving close stellar companions and measuring their relative positions. 

Subsequent systematic speckle programs, including those associated with the CHARA group, provided extensive measurements of binary-star separations and position angles, enabling relative astrometry, orbital characterization, and, in suitable systems, the determination of stellar masses \citep{mcalister1987future, hartkopf1989binary, hartkopf1996binary, mason1999speckle, evans2024orbit}. Long-term programs at the Special Astrophysical Observatory (SAO) and the United States Naval Observatory (USNO) further demonstrated the value of speckle interferometry for monitoring visual binaries and characterizing their orbital motion \citep{balega2002speckle, balega2002vizier, mason2004speckle}. More recently, large speckle surveys, such as those conducted with the SOAR telescope \citep{tokovinin2018ten, tokovinin2024speckle}, have extended these efforts to the systematic characterization of 22 binary star systems \citep{Docobo_2026}. The combination of speckle interferometry with precise astrometric measurements from the Gaia mission further offers an opportunity to improve binary-star orbital solutions and determine fundamental stellar parameters \citep{mitrofanova2024long}. 

SI has also been applied to other science cases, including studies of exoplanet host stars \citep{howell2011} and investigations of stellar surfaces and winds in massive and evolved stars, such as OB-type and Wolf-Rayet stars \citep{weigelt1986eta, lortet1987speckle}. In addition to its primary role in recovering diffraction-limited information on stellar sources, sequences of short-exposure (speckle) images inherently carry statistical signatures of atmospheric turbulence. By analyzing the temporal and spatial statistics of speckle motion and intensity fluctuations, parameters such as atmospheric seeing (Fried parameter), coherence time, and tilt-anisoplanatic angle can be estimated from the data itself, providing valuable site characterization without dedicated instruments \citep{Fried1966, vonDerLuhe1984, Hickson2019Turbulence, Reddy2019Atmospheric}.

This pilot study represents the first attempt to reach the diffraction limit of the 1.3-m Devasthal Fast Optical Telescope (DFOT) using sCMOS-based speckle observations. Despite frame-to-frame positional variations in speckle patterns, we successfully recover the angular relative separation and corresponding projected position angle of components in resolved wide binary systems using Fourier transform techniques, thereby establishing the feasibility of SI with DFOT given the instrument's limited capabilities.  These results motivate dedicated optimization of instruments and detectors, with the long-term goal of enabling routine intermediate-resolution optical interferometry at a meter-class telescope for a broad range of high-impact science cases involving bright stellar objects.

This paper is organized as follows. In the next section~\ref{sec:instr}, we discuss the instrumentation and observational setup of SI at DFOT. Section~\ref{sec:obs} describes the speckle observations of binaries using DFOT. The subsequent section~\ref{sec:res} details the data acquisition strategy and results. We then present a discussion and conclusions that emphasize the role of SI as a practical and powerful intermediate-resolution technique in optical astronomy.
\section{The Telescope and Detector Camera}\label{sec:instr}
\begin{figure*}
\centering
\subfigure[][]{\includegraphics[width=0.43\linewidth]{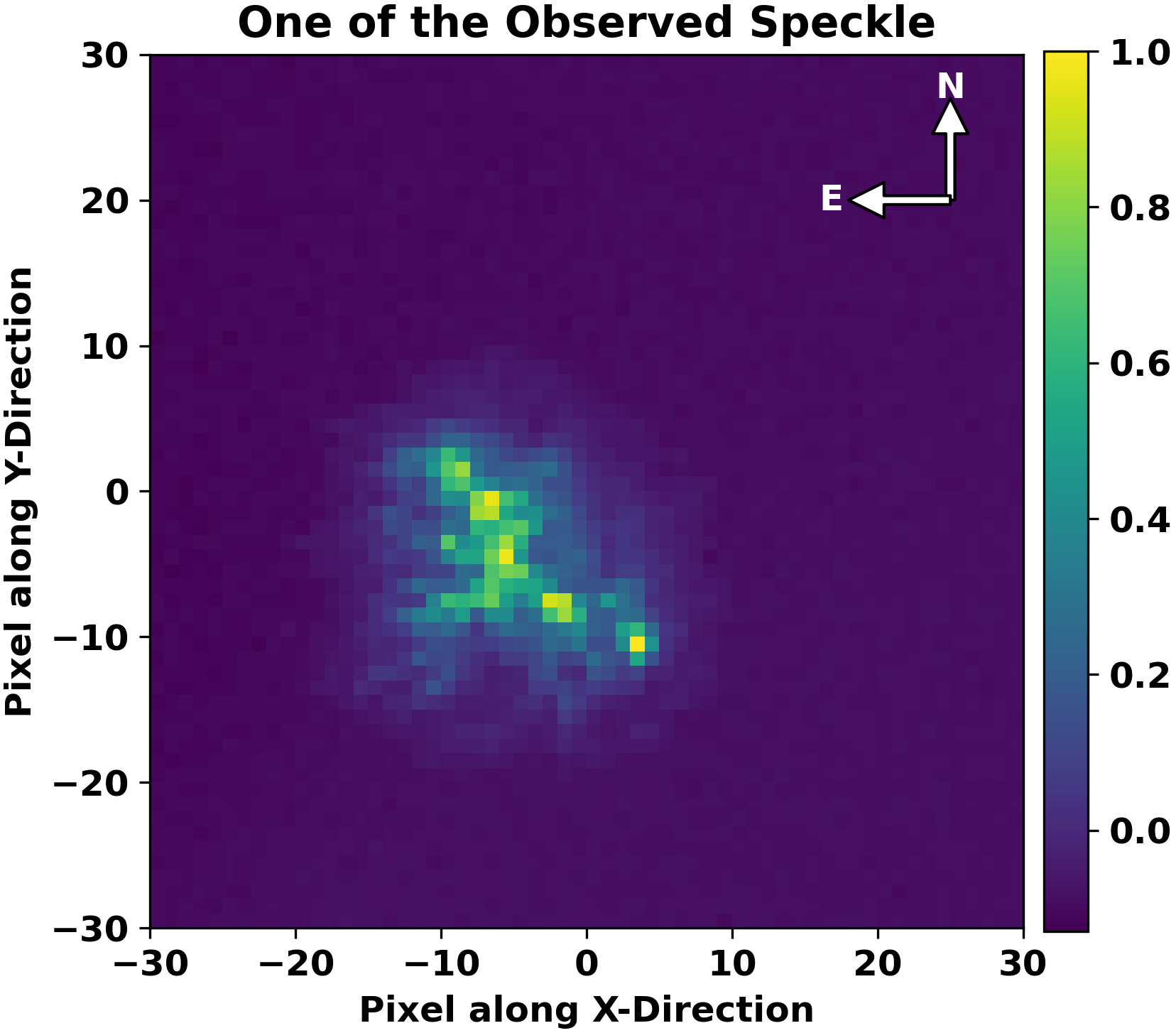}}
\subfigure[][]{\includegraphics[width=0.43\linewidth]{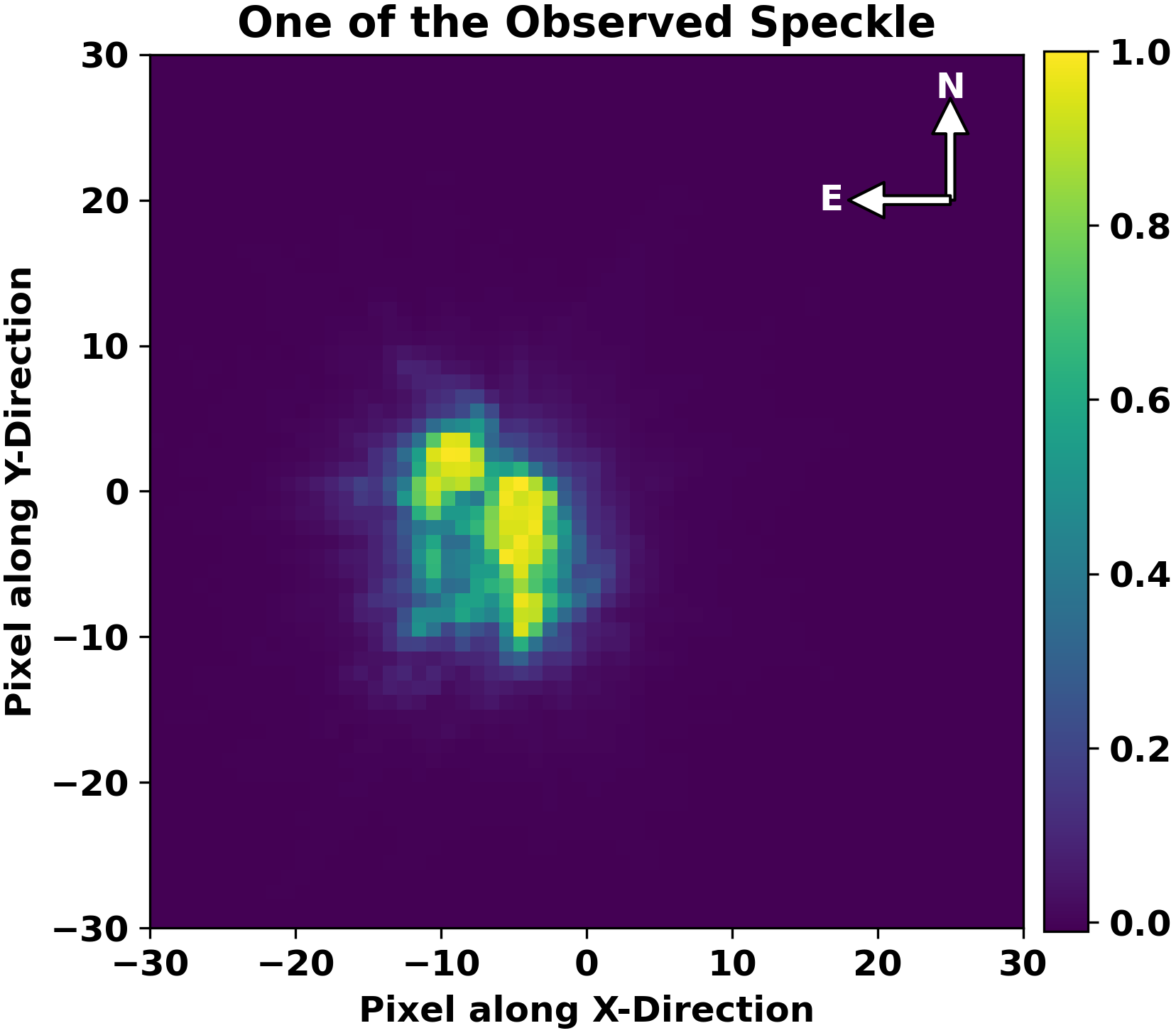}}
\subfigure[][]{\includegraphics[width=0.43\linewidth]{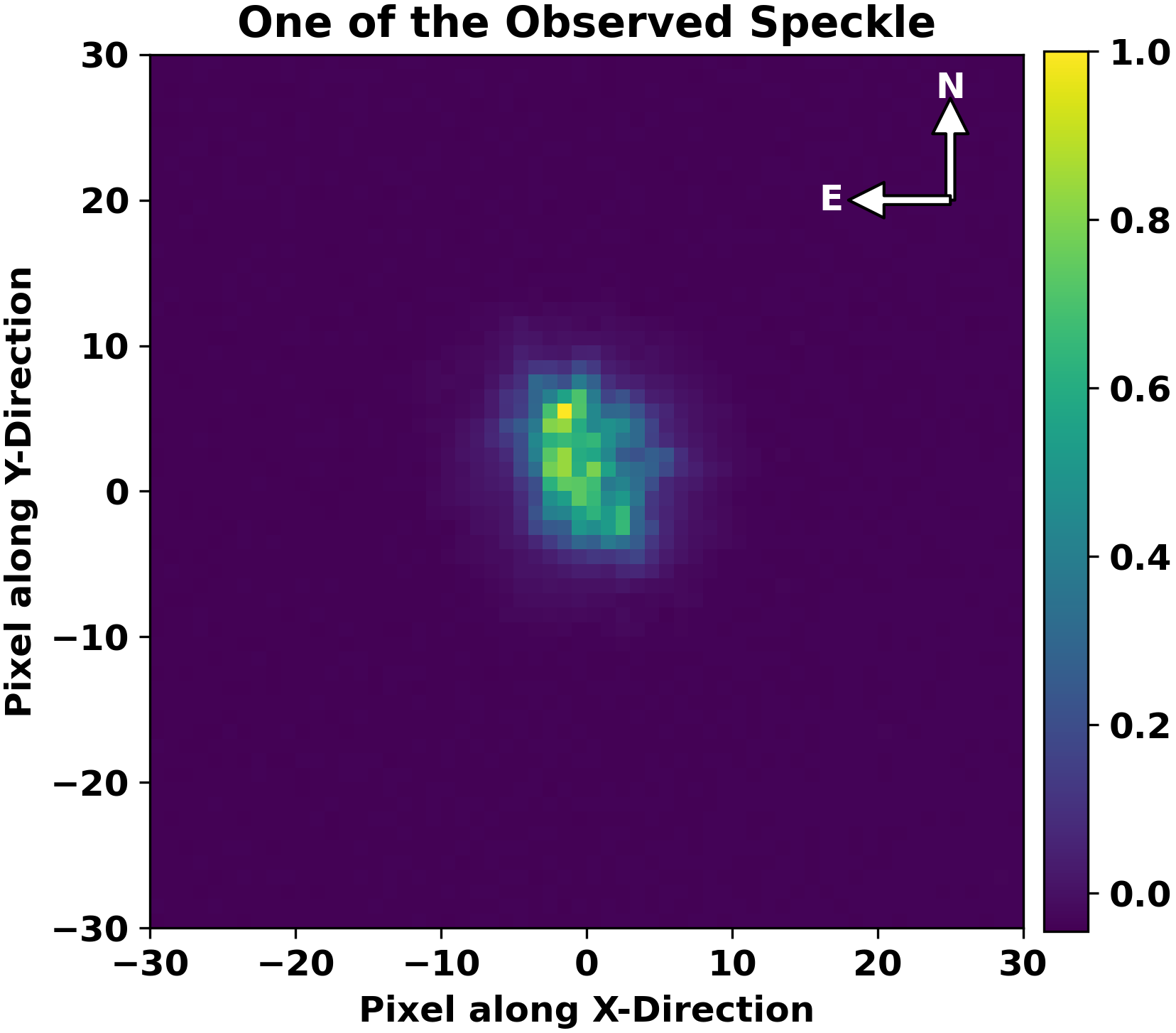}}
\subfigure[][]{\includegraphics[width=0.43\linewidth]{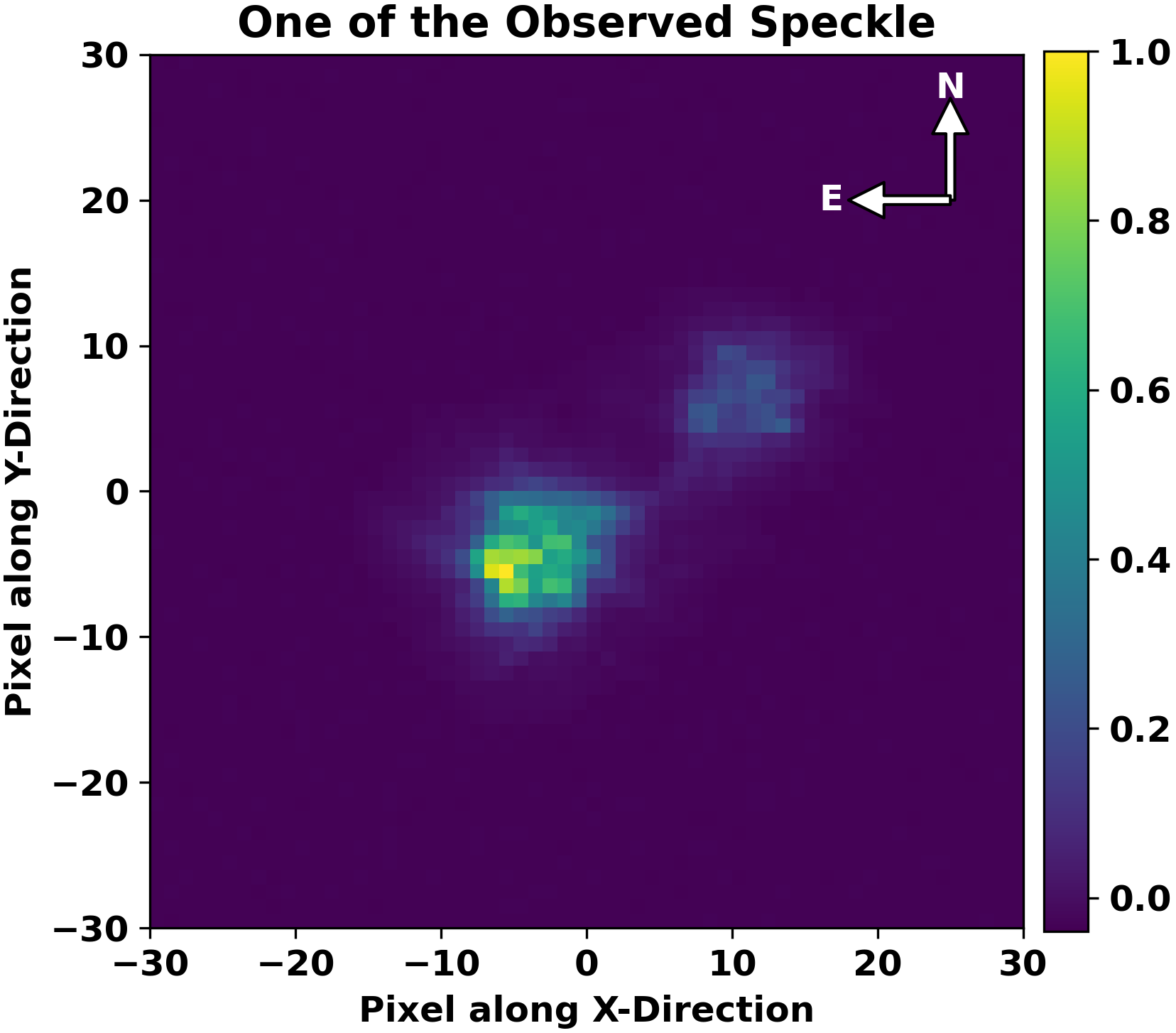}}
\subfigure[][]{\includegraphics[width=0.43\linewidth]{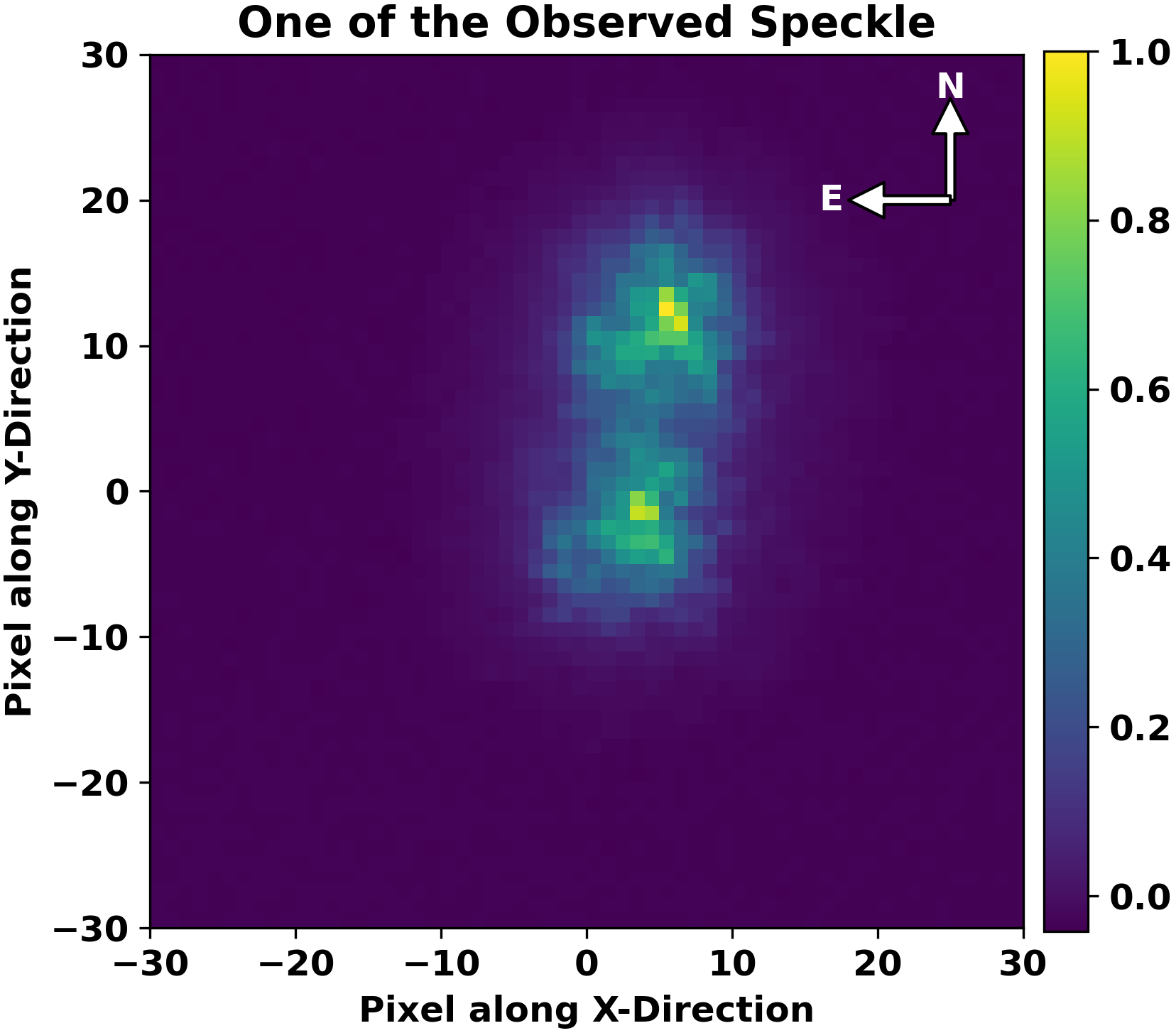}}
\subfigure[][]{\includegraphics[width=0.43\linewidth]{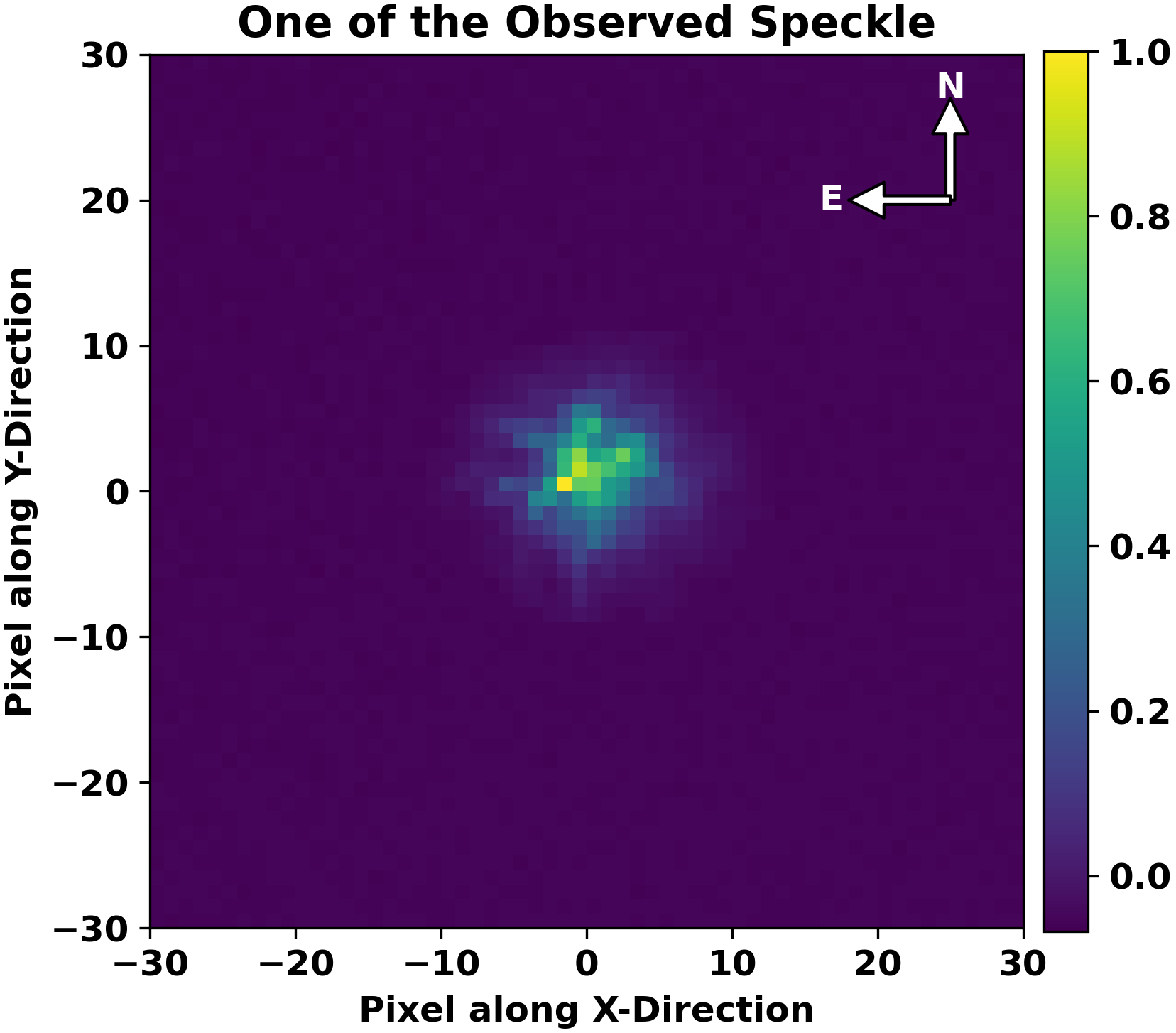}}
\caption{Representative speckle images of six binary systems: (a) $\alpha$~Oph, (b) $\alpha$~Vir, (c) A~Boo, (d) $\gamma$~Leo, (e) $\gamma$~Vir, and (f) $\zeta$~Her. These images correspond to the region of interest extracted from the raw images. The color bars indicate the normalized intensity distribution over the pixels.}
\label{fig:Speck}
\end{figure*}

\begin{figure*}
\centering
\subfigure[][]{\includegraphics[width=0.43\linewidth]{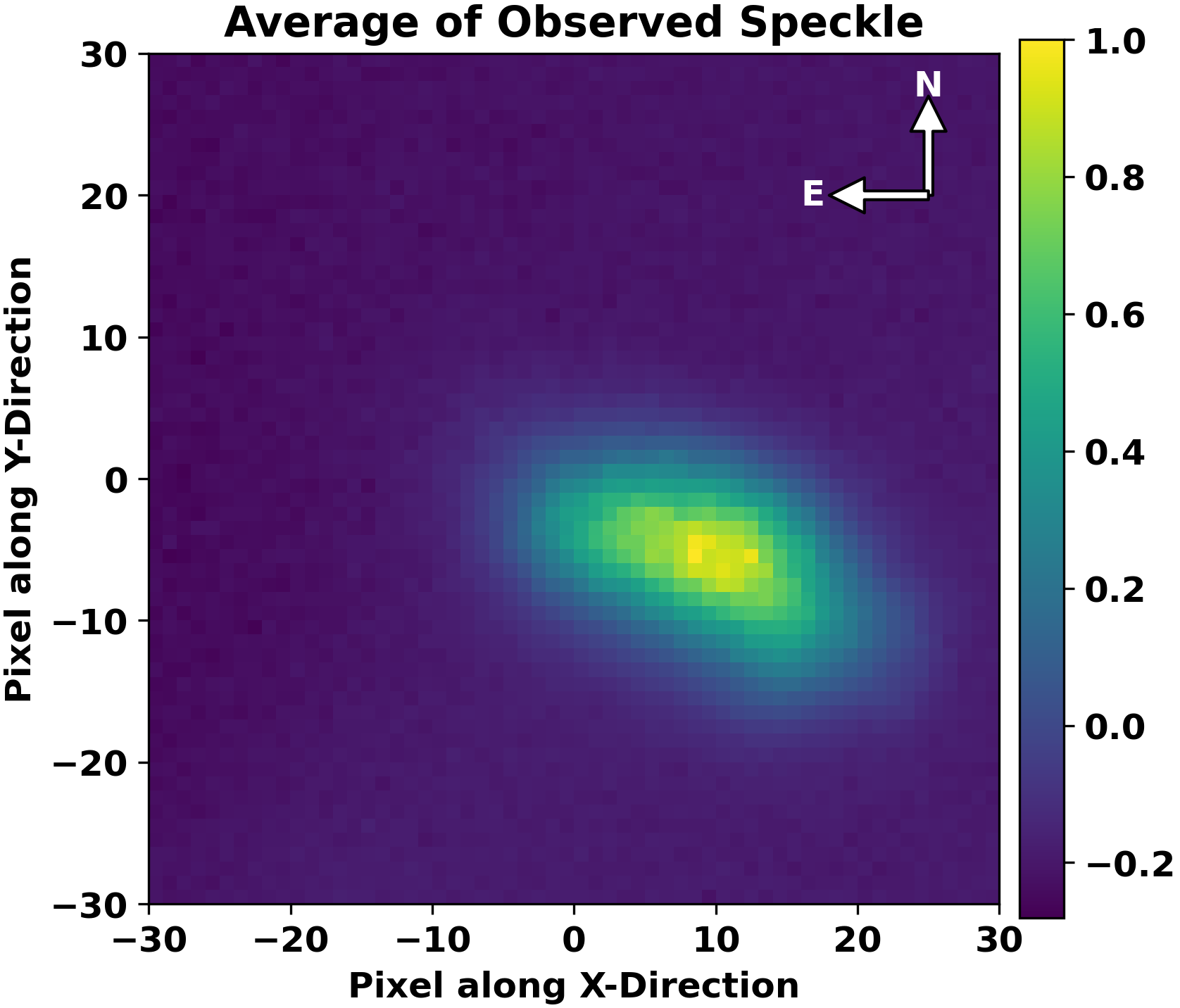}}
\subfigure[][]{\includegraphics[width=0.43\linewidth]{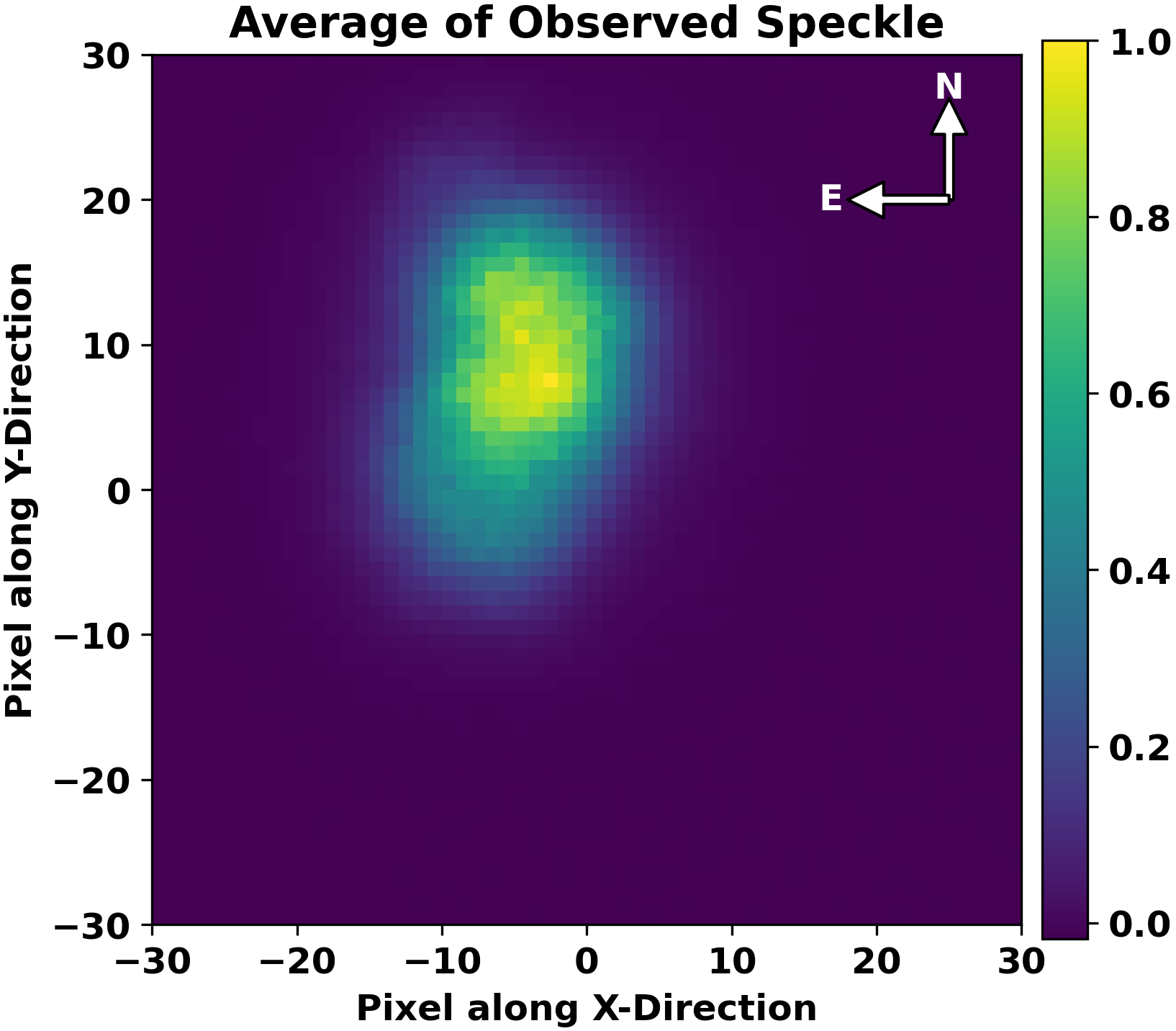}}
\subfigure[][]{\includegraphics[width=0.43\linewidth]{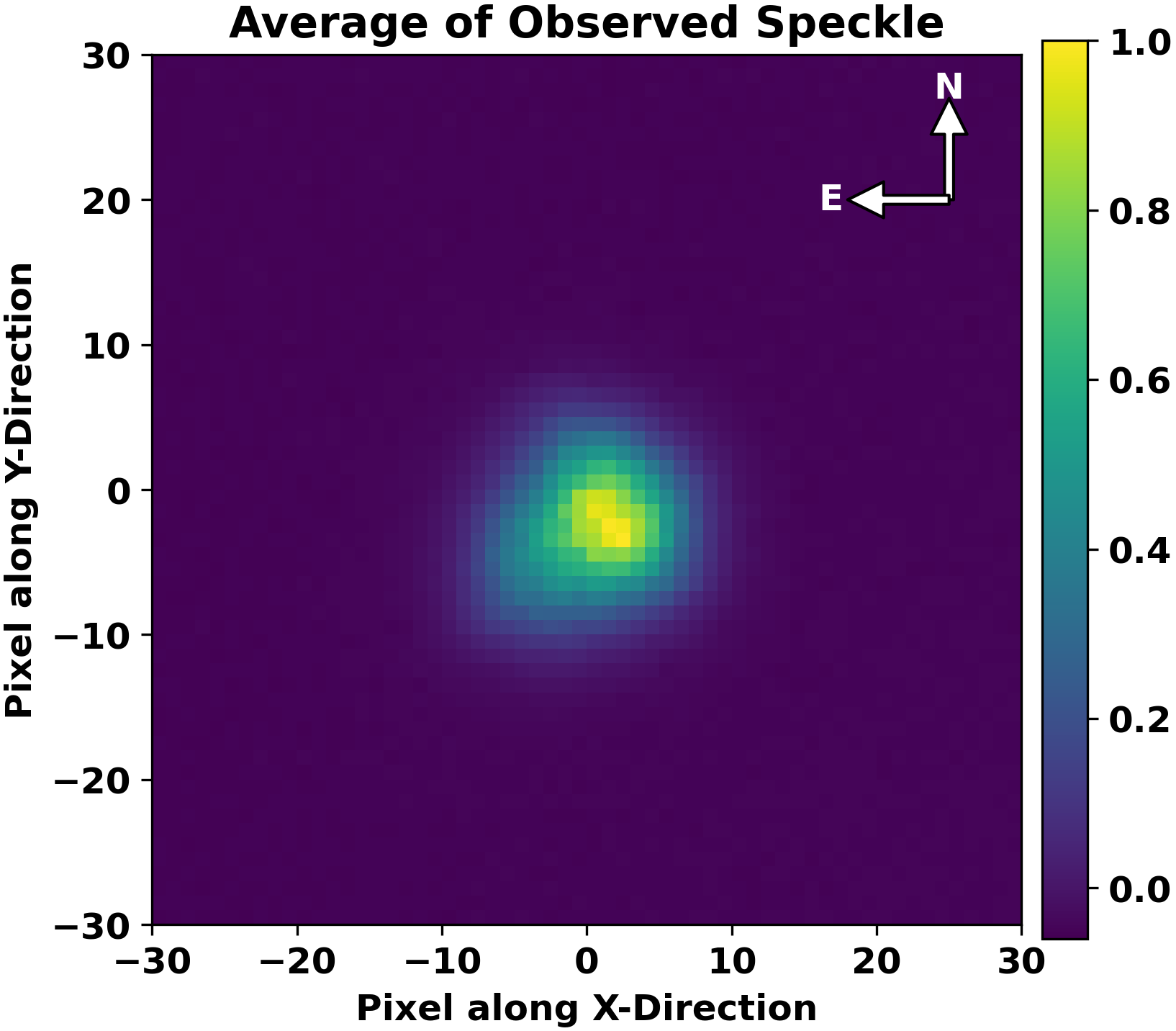}}
\subfigure[][]{\includegraphics[width=0.43\linewidth]{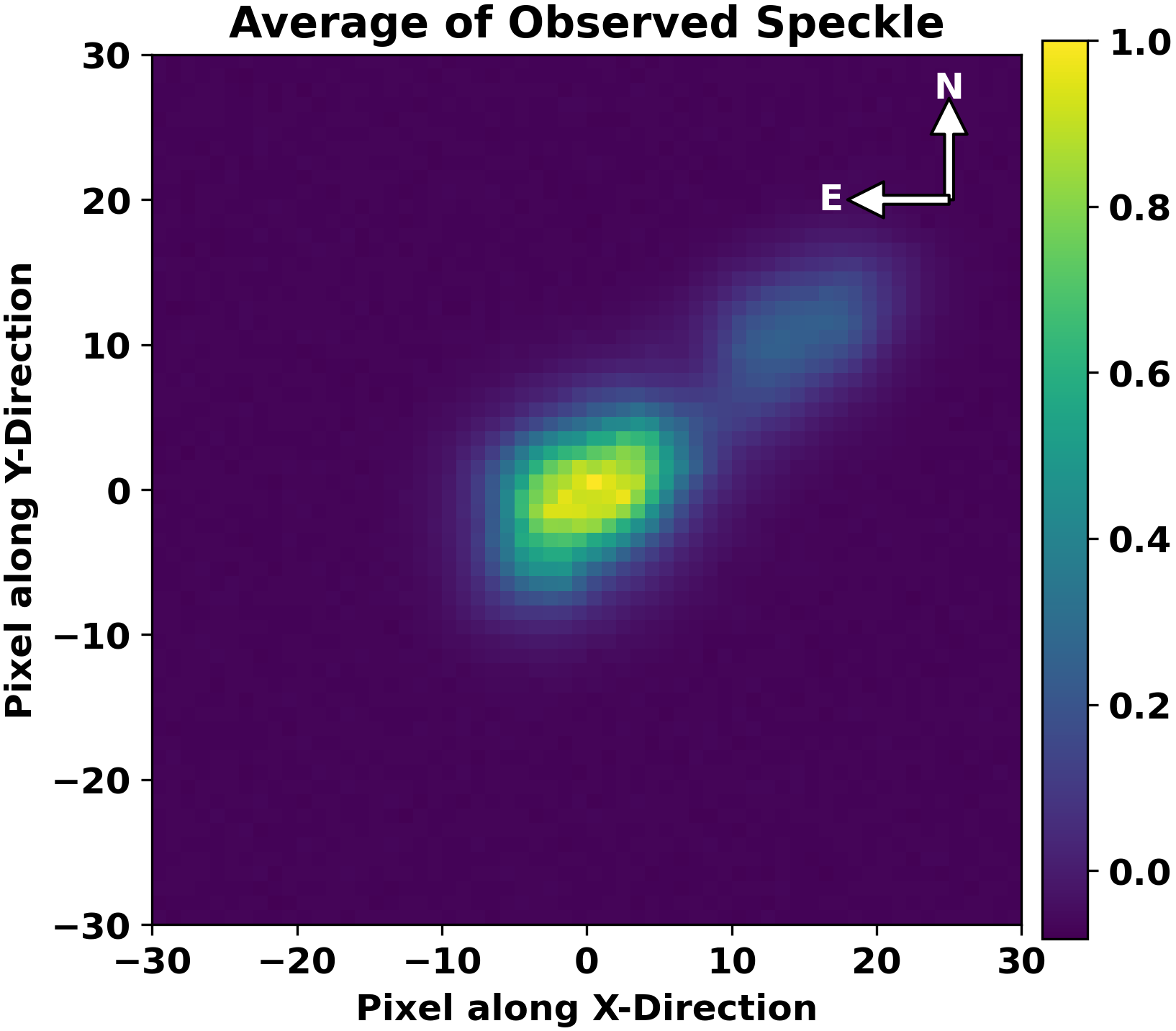}}
\subfigure[][]{\includegraphics[width=0.43\linewidth]{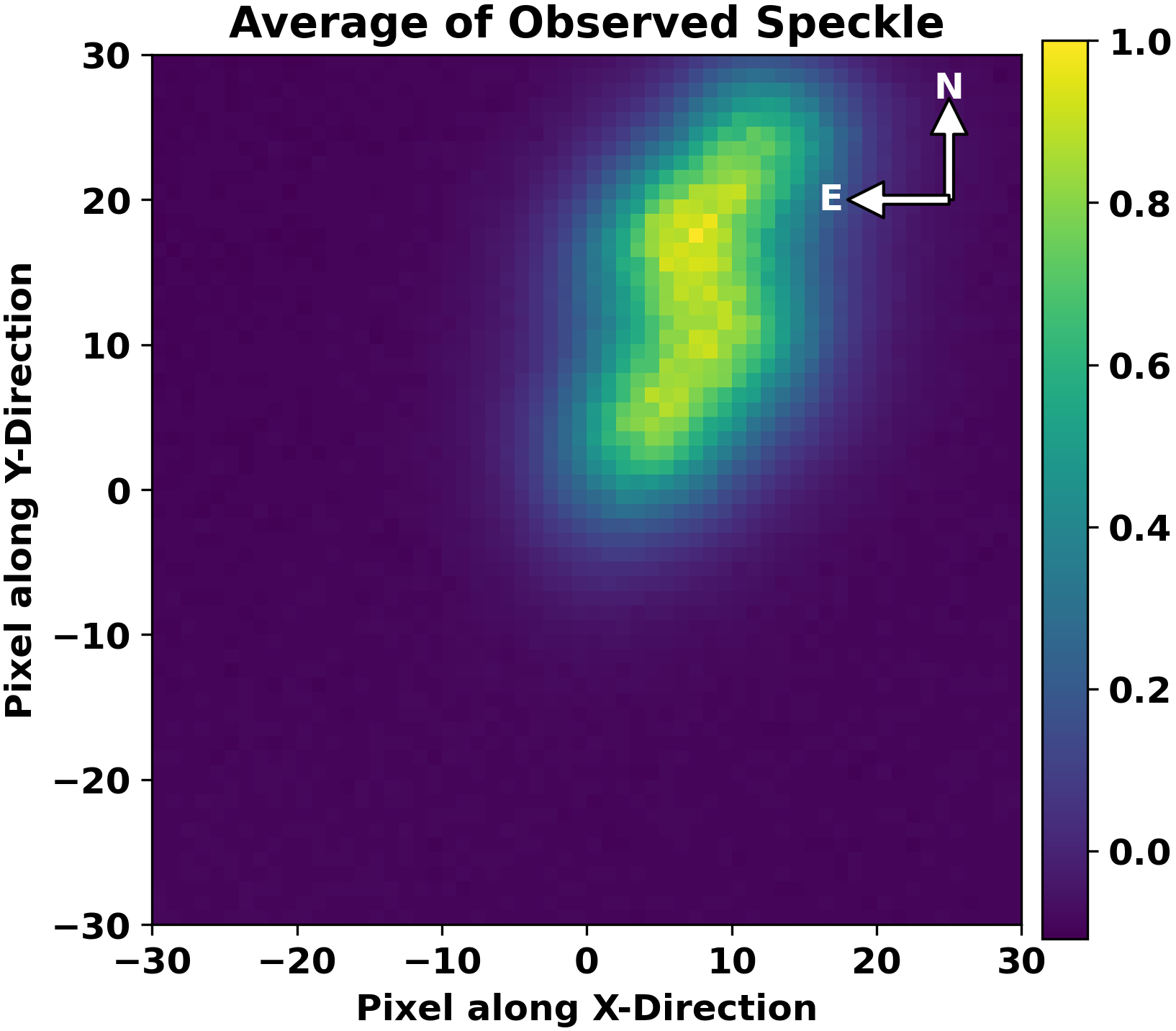}}
\subfigure[][]{\includegraphics[width=0.43\linewidth]{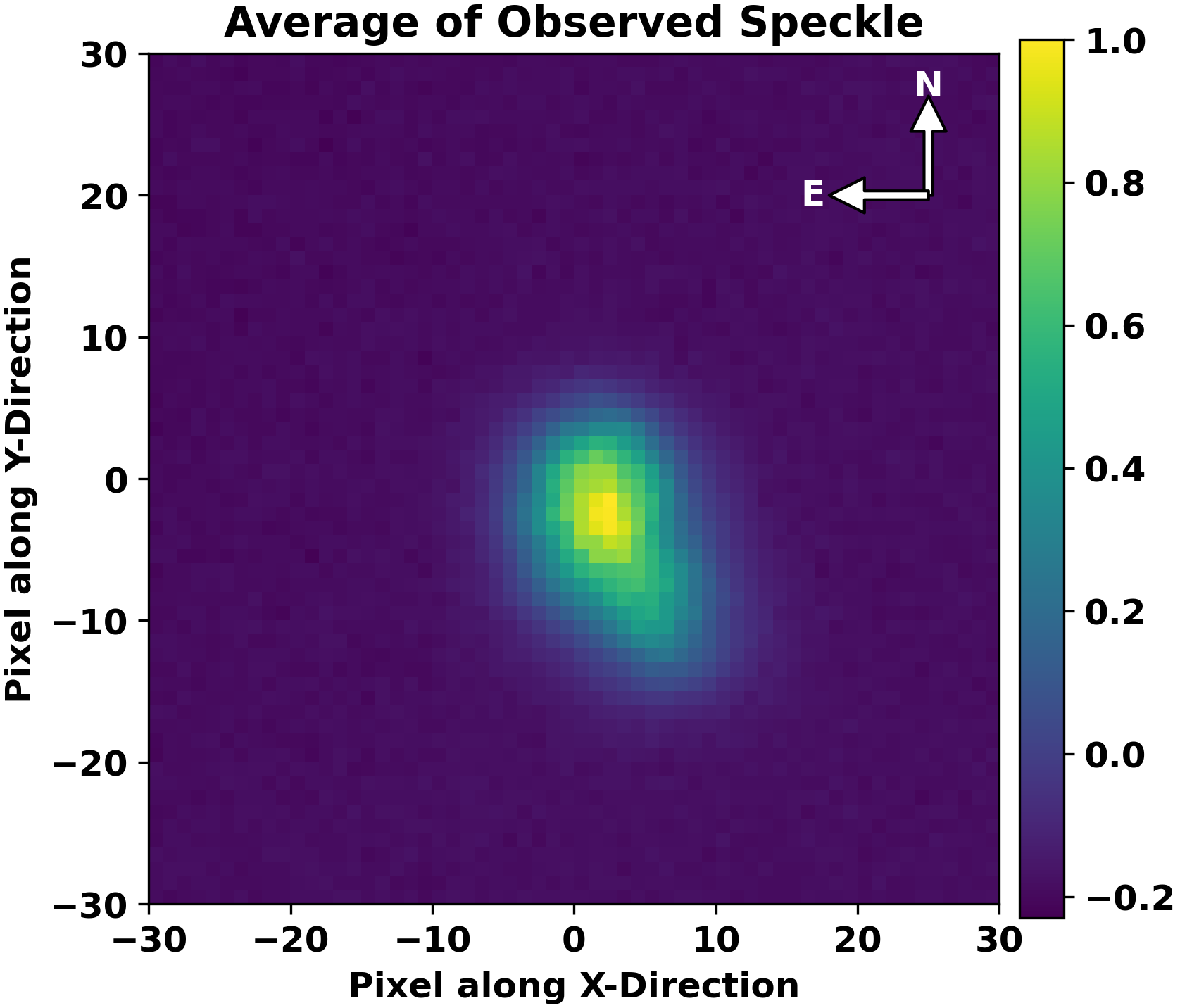}}
\caption{The average of speckle patterns obtained for the selected sample of six binaries: (a) $\alpha$ Oph, (b) $\alpha$ Vir, (c) A Boo, (d) $\gamma$ Leo, (e) $\gamma$ Vir, and (f) $\zeta$ Her. The stretched and deformed nature of each averaged speckle indicates the multi-star constitution of the targets.}
\label{fig:Avg}
\end{figure*}
\begin{table*}[ht!]
\centering
\caption{Summary of the binary systems selected for speckle observations with DFOT. Total Frames represents the number of observed speckle frames. The reference projected separations ($\Delta \theta''_R$) and position angle $\phi_R^\circ$ are taken from the literature, while the DFOT projected separations $\Delta \theta''$ and the corresponding position angle $\phi^\circ$ represent approximate values obtained from the speckle analysis.}
\label{table:binary}

\begin{tabular}{ccccccc}
\toprule
Stars & Frames &
$\sim\Delta\theta_R''$ & $\sim\phi_R^\circ$ &
$\sim\Delta\theta''$ & $\sim\phi^\circ$ &
Reference \\
\midrule

$\alpha$ Oph & 520 & 0.64 & 240 & ... & ... &
\citet{Gardner_2021} \\

$\alpha$ Vir & 300 & 0.001 & 317 & ... & ... &
\citet{pavel2026microarcsecond} \\

A Boo & 220 & 0.019 & 83 & ... & ... &
\citet{198147H,ren2013hipparcos} \\

$\gamma$ Leo & 300 & 4.87 & 128.1 &
$4.744\pm0.012$ & $126.94\pm0.14$ &
\citet{harshaw2015calibrating} \\

$\gamma$ Vir & 500 & 2.45 & 9.8 &
$3.387\pm0.027$ & $171.57\pm0.33$ &
\citet{harshaw2015calibrating} \\

$\zeta$ Her & 300 & 1.35 & 131.3 &
$1.582\pm0.026$ & $81.06\pm0.91$ &
\citet{morel2001zeta} \\

\bottomrule
\end{tabular}
\end{table*}

The 1.3-m DFOT employs a Ritchey-Chretien Cassegrain optical configuration with a relatively fast focal ratio (f/4), optimized for wide-field imaging and time-domain observational programs. The implementation of SI on the DFOT is intended to augment observations conducted with the heavily subscribed 3.6-m Devasthal Optical Telescope (DOT) \citep{rai2025stellar}. The DOT is likewise based on a Ritchey-Chretien Cassegrain design but operates at a slower effective focal ratio of f/9. 

The significantly larger aperture of the DOT results in a smaller diffraction limit and hence superior intrinsic angular resolution. In contrast, the smaller 1.3-m aperture of the DFOT provides a suitable testbed for assessing the practicality and performance of SI with a more modest, meter-class telescope. In particular, DFOT can furnish complementary capabilities for observations of bright binary and multiple stellar systems, for repeated temporal monitoring, and for extensive observational campaigns in which the angular-resolution and sensitivity requirements are adequately satisfied by a meter-class aperture.

Achieving diffraction-limited resolution with any telescope requires spatially well-resolved speckle patterns distributed across multiple pixels on the detector. To satisfy this requirement, we employed an MP Andor Marana sCMOS camera. The camera is mounted at the focal plane of DFOT such that the resultant images are oriented with North (N) upward and East (E) toward the left, as illustrated in Figure~\ref{fig:Speck}. The detector comprises $2048 \times 2048$ pixels, each with a physical size of 6.5~$\mu$m, yielding a total field of view of approximately $8.8 \times 8.8$~arcmin$^{2}$ on the sky. The camera has a gain of $\sim$1 e$^-$/ADU and a readout noise of $\sim$1.6 e$^-$. Since speckle observations do not require a large field of view, only a selected subregion of the detector is used for data acquisition. As shown in Figure~\ref{fig:Speck}, the obtained speckle patterns appear undersampled, indicating that the observations did not reach the diffraction limit and highlighting the need for further instrumental optimization to improve speckle contrast.

The minimum diffraction-limited angular scale for a telescope of diameter $D$, observing at wavelength $\lambda$, is given by
\begin{equation}
    \Delta\theta \approx 1.22\frac{\lambda}{D},
    \label{eqn:diff}
\end{equation}
which corresponds to $0.126''$ for the 1.3-m DFOT at 650~nm. The pixel scale of the sCMOS detector used here is $0.258''~\mathrm{pixel}^{-1}$. For Nyquist sampling, at least two pixels are required per diffraction-limited resolution element (\citealp[e.g.][]{hearnshaw2009astronomical}). The corresponding required pixel scale is therefore
\begin{equation}
    p_{\rm Nyquist}
    =
    \frac{\Delta\theta}{2}
    =
    \frac{0.126''}{2}
    \approx 0.063''~\mathrm{pixel}^{-1}.
\end{equation}
Thus, the present detector sampling of $0.258''~\mathrm{pixel}^{-1}$ is approximately four times coarser than the required Nyquist sampling, and the diffraction-limited speckle structure is therefore significantly undersampled. This undersampling quite naturally contributes to the reduced speckle contrast observed in our data, as shown in Figure~\ref{fig:Speck}. Future observations will require either a detector with a finer effective pixel scale or an appropriate optical magnification system to provide adequate spatial sampling.  An additional optical element, such as a focal extender, along with optimized detector performance, is necessary to achieve proper sampling.
\section{Observations of Speckles}\label{sec:obs}
Speckle images of six binary stars namely $\alpha$~Oph, $\alpha$~Vir,  A~Boo, $\gamma$~Leo, $\gamma$~Vir, and $\zeta$~Her were taken on 5th March 2025 in the R-band (effective wavelength $\sim$ 650 nm).  Details of the observed targets are provided in Table~\ref{table:binary}. Each speckle frame was acquired with an exposure time of 2~ms. 
For each target, 20 frames were recorded per set, with a total of 26 sets for $\alpha$~Oph, 15 sets for $\alpha$~Vir, 11 sets for A~Boo, 15 sets for $\gamma$~Leo, 25 sets for $\gamma$~Vir, and 15 sets for $\zeta$~Her.
Several bias and twilight flat frames were also taken during the observations. 
All images were pre-processed to minimize instrumental systematics using standard bias subtraction and flat-field normalization procedures before subsequent analysis. 

Figure~\ref{fig:Speck} shows representative speckle patterns for each observed target. The present analysis focuses on detecting and measuring the projected angular separation and position angle of the observed binary components, and thus does not require photometric calibration. 
Determining the apparent magnitudes of the binary components was beyond the scope of this study; therefore,  the lack of standard calibration star observations does not impact the results presented in this work.

The instantaneous speckle images taken over time do not provide stable spatial information due to strong atmospheric turbulence. For wide binary systems, orbital signatures can be identified directly in individual speckle frames, but the associated positional uncertainties remain large owing to the rapid temporal evolution of speckle patterns driven by atmospheric seeing variations. In the case of close binary systems, the speckle patterns are neither sufficiently well resolved nor clear enough to achieve diffraction-limited performance with the current instrumental configuration. Furthermore, averaging over all recorded frames does not yield a static or clearly resolved orbit, as illustrated in Figure~\ref{fig:Avg}. However, the generally stretched and tailed structure of these averaged images, and in at least three cases, $\gamma$ Leo, $\gamma$ Vir, and $\zeta$ Her, the distinctly visible centers of light emission, suggest a multi-star composition for these targets.
\section{Speckle-Induced Fluctuations in Orbit}\label{sec:res}
\begin{figure*}
\centering
\subfigure[][]{\includegraphics[width=0.43\linewidth]{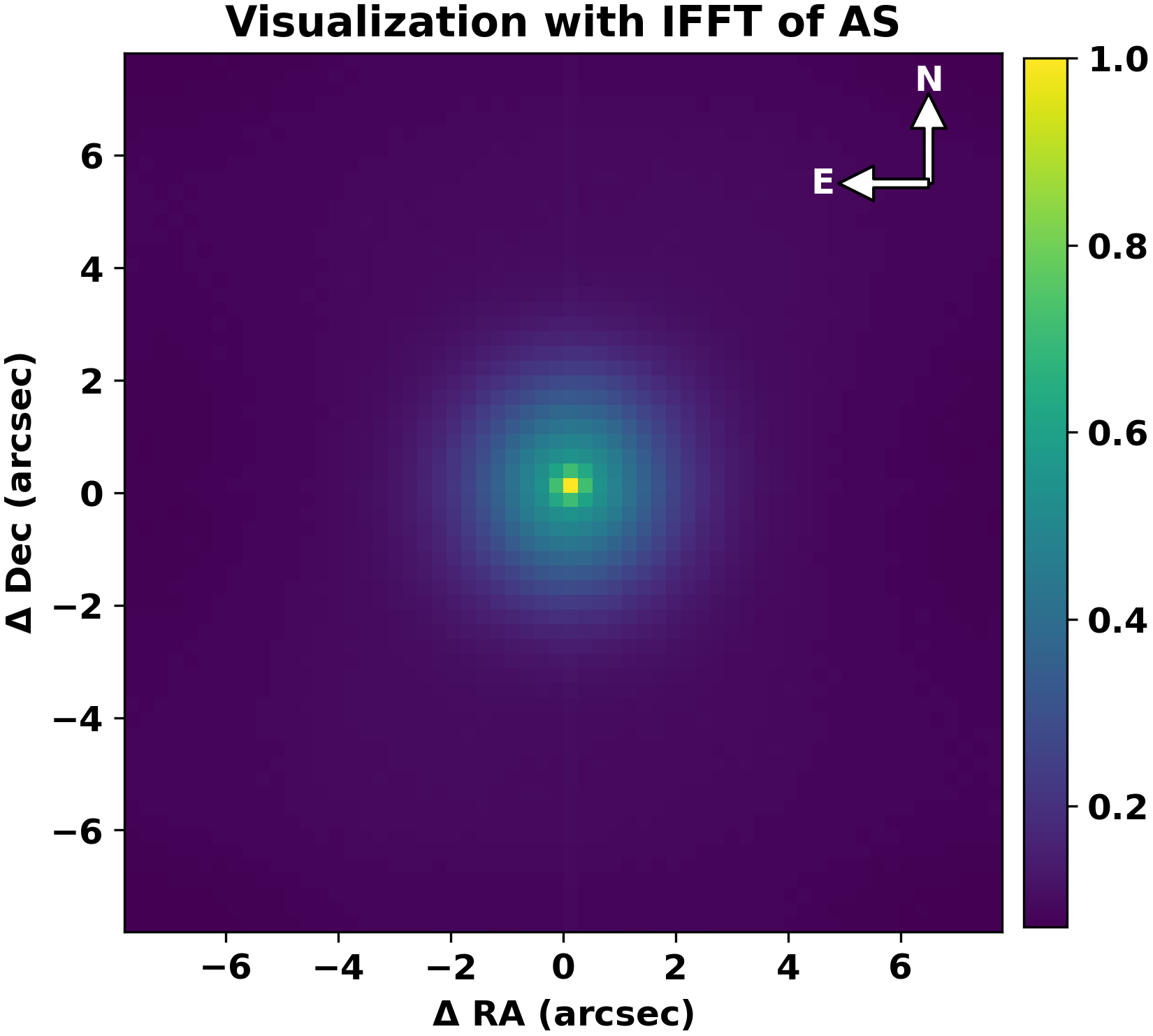}}
\subfigure[][]{\includegraphics[width=0.43\linewidth]{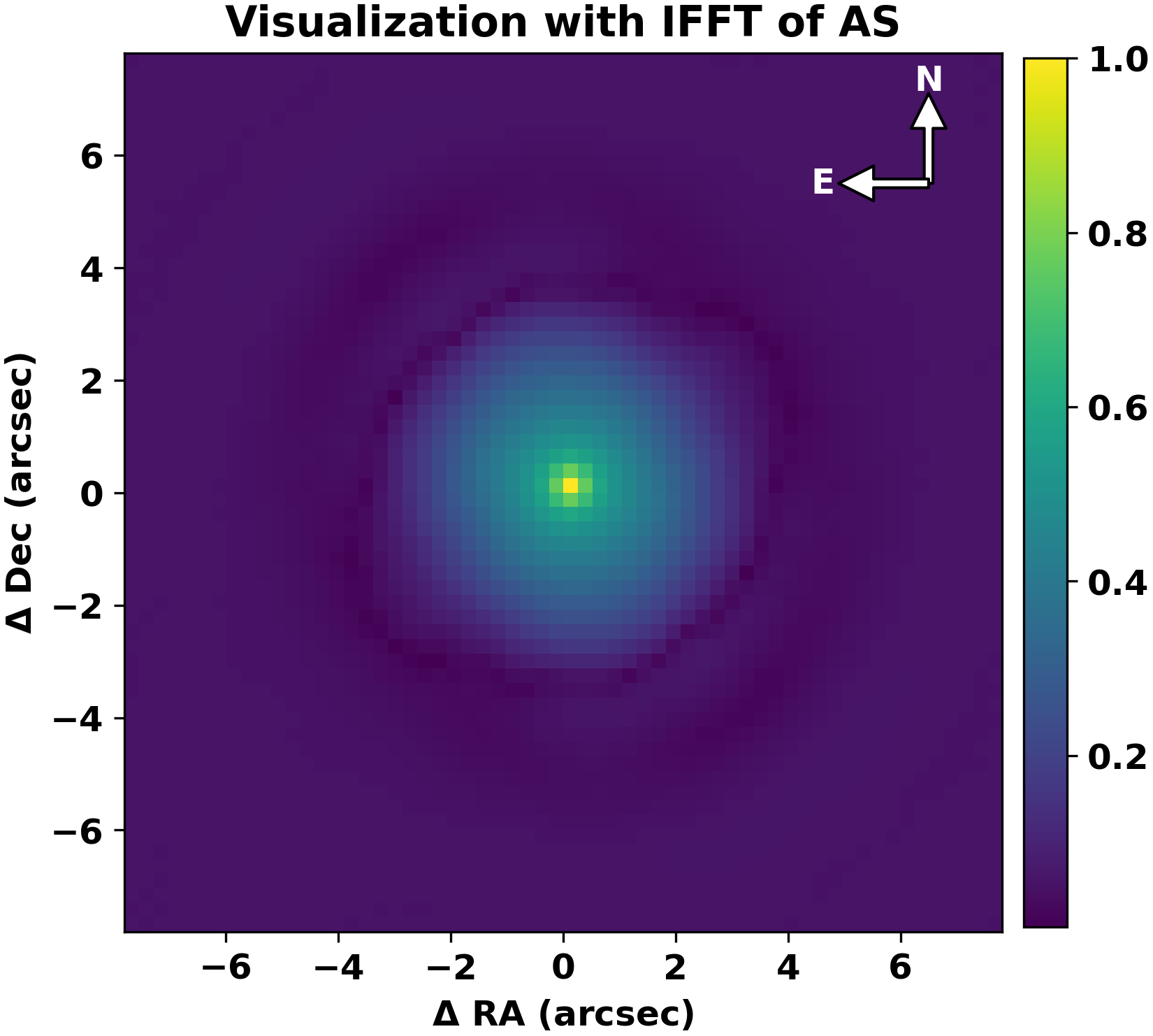}}
\subfigure[][]{\includegraphics[width=0.43\linewidth]{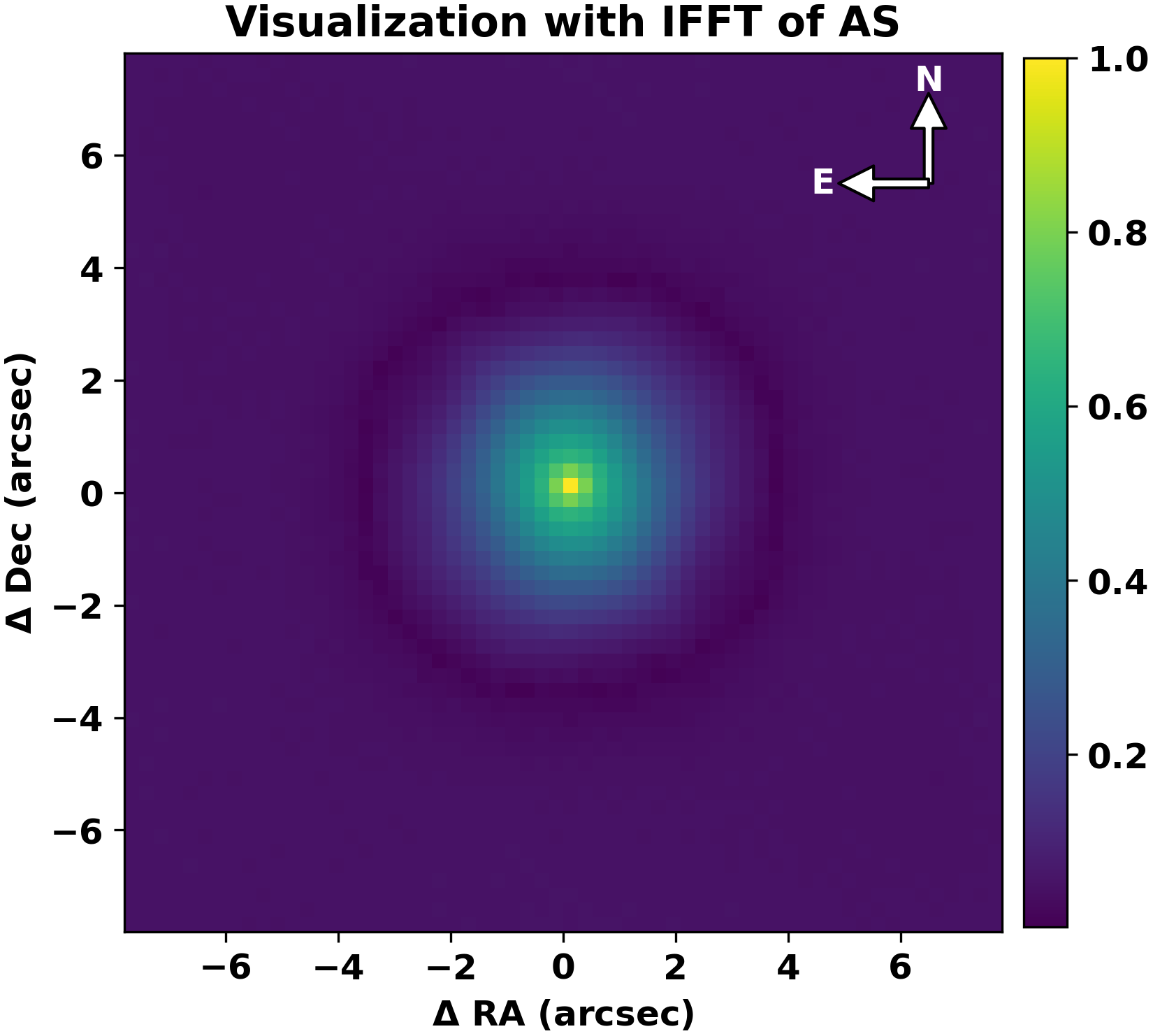}}
\subfigure[][]{\includegraphics[width=0.43\linewidth]{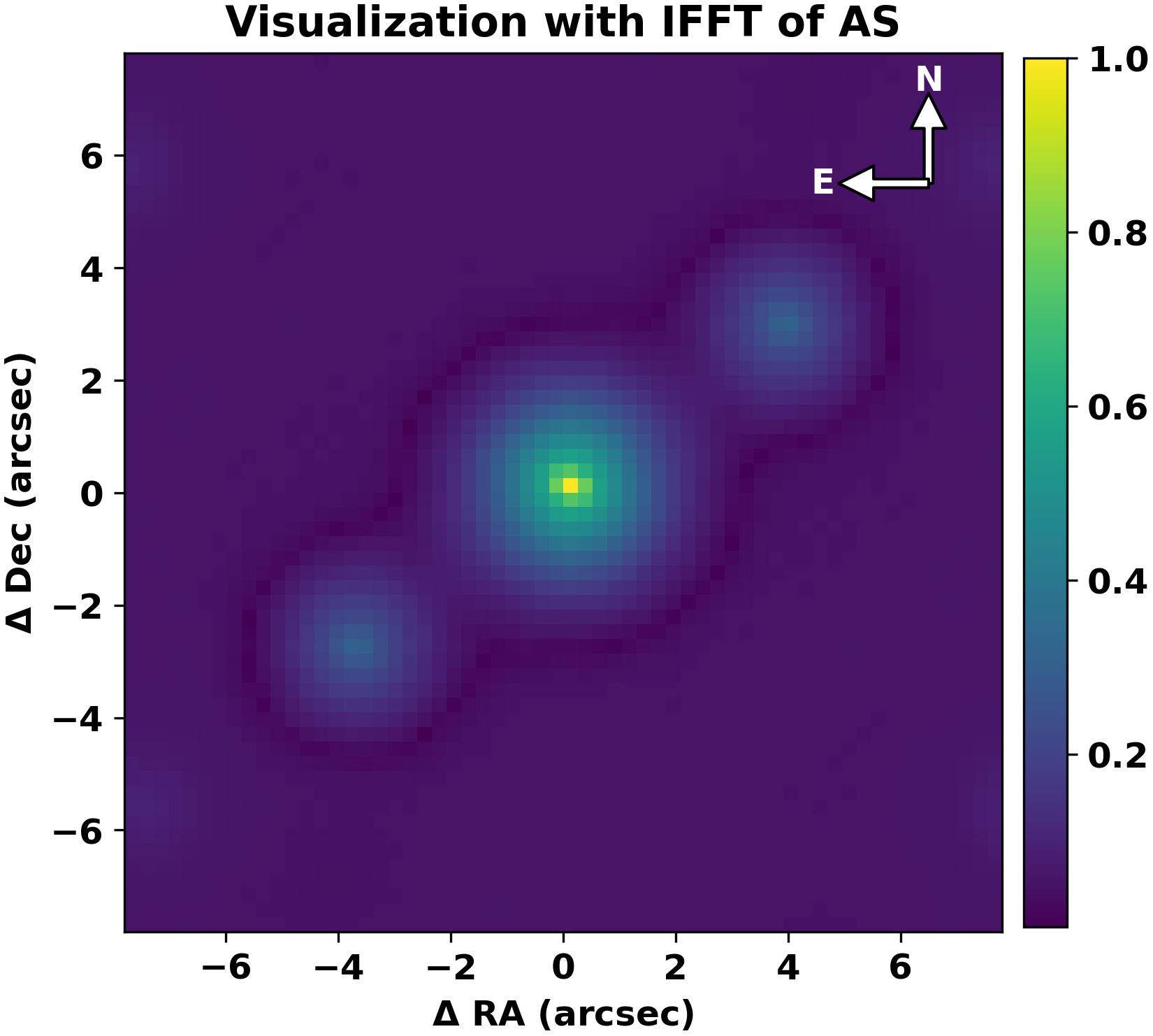}}
\subfigure[][]{\includegraphics[width=0.43\linewidth]{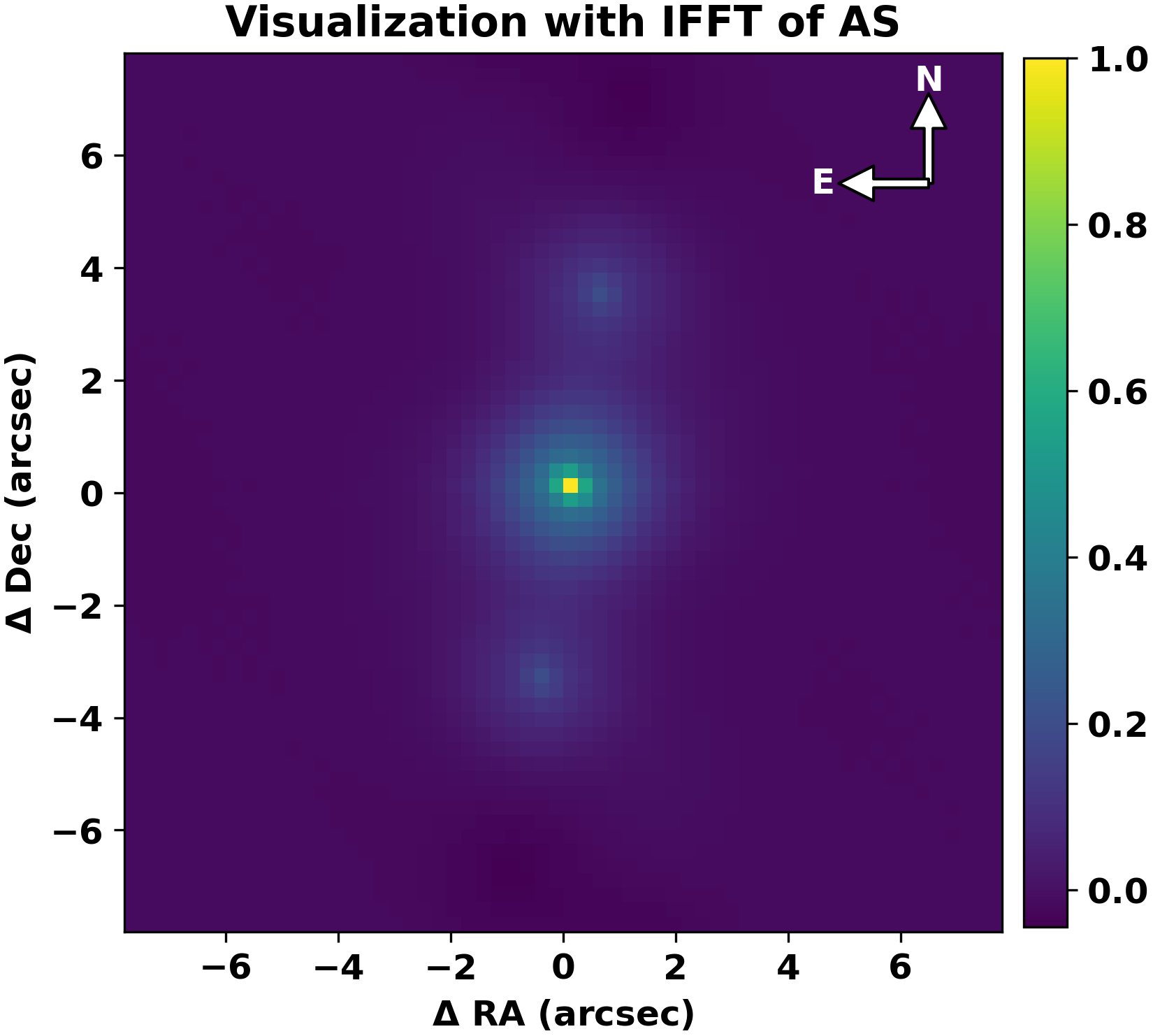}}
\subfigure[][]{\includegraphics[width=0.43\linewidth]{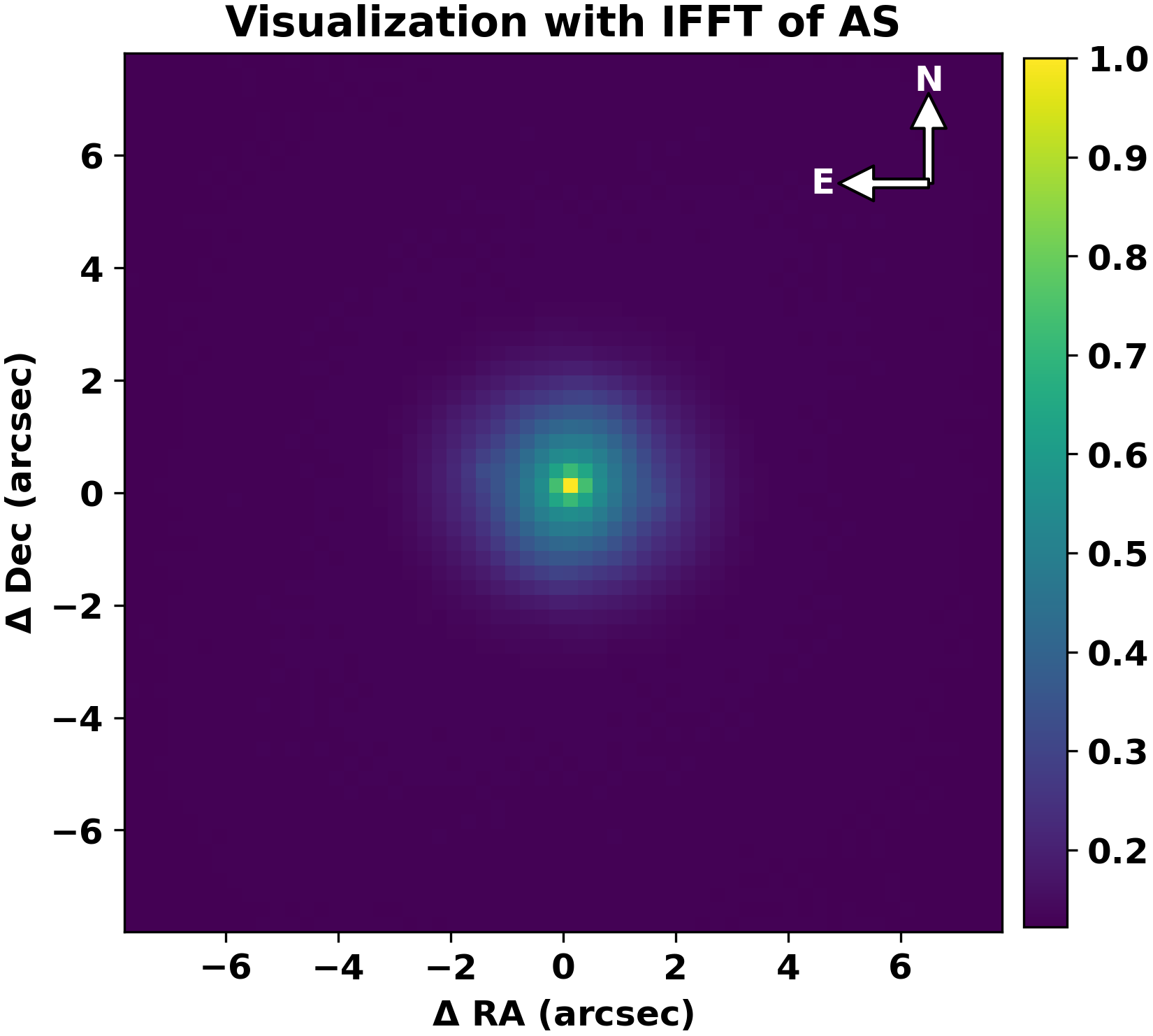}}
\caption{Inverse Fast Fourier Transform of the amplitude spectrum of speckle patterns averaged over several frames for the six binary systems (a) $\alpha$ Oph, (b) $\alpha$ Vir, (c) A Boo, (d) $\gamma$ Leo, (e) $\gamma$ Vir, and (f) $\zeta$ Her. The sidelobe nature of $\alpha$ Vir and A Boo, the secondary peaks in $\gamma$ Leo and $\gamma$ Vir, and the faint secondary peaks in $\zeta$ Her are clearly discernible.}
\label{fig:Auto}
\end{figure*}

The recorded speckle frames as shown in Fig.~\ref{fig:Speck} were analyzed using a Fourier-domain approach. For each bias- and flat-field-corrected frame, the Fourier transform was computed, and the corresponding power spectrum was obtained. The power spectra obtained from the individual frames were then averaged to improve the signal-to-noise ratio. In the process of this averaging,  the essentially random effects of atmospheric seeing are minimized. In the large-number limit of image frames of a particular binary system, these effects should cancel out. The inverse Fast Fourier transform (IFFT) of the absolute value of the averaged power spectrum was subsequently computed to obtain the corresponding autocorrelation-like distribution, as shown in Fig.~\ref{fig:Auto}. Square-root scaling is applied uniformly to each pixel value, ensuring that even the secondary dots are clearly visible.

As mentioned in section~\ref{sec:instr} using eqn.~\ref{eqn:diff}, the minimum angular resolution achievable with DFOT using the SI technique is $0.126''$ at a wavelength of 650~nm. The two selected close binaries, $\alpha$ Vir ($\Delta \theta\approx 0.001''$; \citealp{pavel2026microarcsecond}) and A~Boo ($\Delta \theta\approx 0.019''$; \citealp{198147H}), are not possible to resolve even with updated instrumentation at the backend of DFOT, as shown in second and third panel from top left to bottom right of Fig.~\ref{fig:Auto}. In contrast, the intermediate-separation binary $\alpha$~Oph ($\Delta \theta\approx 0.64''$; \citealp{Gardner_2021}), which remains unresolved in the present observational attempt (top left panel of Fig.~\ref{fig:Auto}), is expected to be resolved using the SI technique with the upgraded instrumentation discussed in the final paragraph of Section~\ref{sec:instr}.

For the wider binary systems, the Fourier analysis reveals characteristic side-lobe structures associated with binary components. These features are clearly visible for the binary $\gamma$ Leo, $\gamma$ Vir, and $\zeta$ Her as shown in fourth, fifth and sixth panel from top left to bottom right of Figs. ~\ref{fig:Auto}. This demonstrates that the relative geometry of the wider components can be recovered from the recorded speckle data. The projected relative separations obtained for $\gamma$~Leo, $\gamma$~Vir, and $\zeta$~Her are approximately $4.744'' \pm 0.012''$, $3.387'' \pm 0.027''$, and $1.582'' \pm 0.026''$, respectively, with corresponding projected position angles of $126.94^{\circ} \pm 0.14^\circ$, $171.57^{\circ} \pm 0.33^\circ$, and $81.06^{\circ} \pm 0.91^\circ$, respectively. 

These results have also been compared with the previously measured value, and Table~\ref{table:binary} summarizes them along with the literature values and their references. The measured separation and position angle are affected by frame-to-frame variations in the observed speckle patterns, primarily due to atmospheric turbulence and the resulting temporal variations in the speckle distribution. Averaging the Fourier-domain information over multiple frames reduces these temporal fluctuations and enhances the characteristic binary signatures. Nevertheless, the measured parameters should be regarded as preliminary because the present observations were obtained without sufficiently resolved diffraction-limited speckle patterns. Overall, the results demonstrate that the present DFOT configuration can recover the signatures of relatively wide binary systems, while the detection and characterization of closer systems require improved spatial sampling, speckle contrast, and instrumental performance.

\section{Summary} 
We have conducted a feasibility study of speckle interferometry using a $2048\times2048$ sCMOS camera mounted on the 1.3-m DFOT. Short-exposure speckle patterns for several binary systems were successfully recorded, yielding their corresponding Fourier-domain power spectra and autocorrelation-like distributions. For relatively wide binary systems, characteristic side-lobe structures were recovered, allowing estimates of their projected separations and position angles. However, the intermediate binary systems remained unresolved because the recorded speckle patterns lacked sufficient spatial resolution and contrast to achieve the diffraction-limited performance of DFOT. In particular, the present detector configuration provides insufficient spatial sampling of the diffraction-limited speckle structure, highlighting an important instrumental limitation of the current setup.

Despite these limitations, the observations demonstrate the feasibility of recording and analyzing speckle patterns with an sCMOS detector on a meter-class telescope. The results provide a useful assessment of the current instrumental limitations and identify the requirements for achieving diffraction-limited speckle observations with DFOT. 
Incorporating a focal extender alongside a suitable detector offers a viable approach to enhancing spatial sampling in future observations. Such improvements would enable more reliable reconstruction of intermediate binary systems and broaden the potential application of speckle interferometry with DFOT to binary and multiple-star characterization, repeated monitoring, and other high-angular-resolution stellar studies.

\section*{Acknowledgments}
We sincerely thank the anonymous referee for the constructive comments and valuable suggestions, which have significantly improved the quality of this manuscript. The authors thank the engineers and technical staff at the observatory site for their invaluable support during the observations and instrumentation efforts. SS thanks the hospitality and the efficient availability of computational facilities at IUCAA, Pune, under its flagship Visiting Associate Programme.

\bibliographystyle{aa}
\bibliography{main}
\end{document}